\documentclass[a4paper]{article}
\usepackage{jheppub}
\usepackage[utf8x]{inputenc}
\usepackage{color}
\usepackage{amsmath}
\usepackage{amssymb}
\allowdisplaybreaks 
\usepackage{xargs}
\usepackage{tikz}
\usepackage{dcolumn}
\usepackage{array}
\usepackage{caption}
\usepackage{pbox}
\usepackage{float}
\usepackage{graphicx}
\usepackage[all]{xy}
\usepackage{longtable}

\numberwithin{equation}{section}

\begin{document}

\thispagestyle{empty}

\rightline{\small}

\vskip 3cm
\noindent
\noindent
{\LARGE \bf Counting the Massive Vacua of  ${\cal N}=1^\ast$ Super
}
\vskip .4cm
\noindent
{\LARGE \bf   Yang-Mills Theory}
\vskip .3cm
\begin{center}
\linethickness{.06cm}
\line(1,0){440}
\end{center}

\vskip .8cm

\noindent
{\large \bf Antoine Bourget and Jan Troost}
\vskip 0.2cm
{\em \hskip -.05cm Laboratoire de Physique Th\'eorique\footnote{Unit\'e Mixte du
CNRS et
    de l'Ecole Normale Sup\'erieure associ\'ee \`a l'universit\'e Pierre et
    Marie Curie 6, UMR
    8549.}}
    \vskip -.05cm
{\em \hskip -.05cm Ecole Normale Sup\'erieure}
 \vskip -.05cm
{\em \hskip -.05cm 24 rue Lhomond, 75005 Paris, France}

\vskip 1cm

\vskip 0.6 cm

\noindent {\sc Abstract :}

\vskip  0.15cm

\noindent

We compute the number of massive vacua  of ${\cal N}=4$ supersymmetric Yang-Mills theory on $\mathbb{R}^4$
 mass-deformed to preserve ${\cal N}=1$ supersymmetry, for any gauge group $G$. We use semi-classical techniques and efficiently reproduce the known
counting for $A,B$ and $C$ type gauge groups, present the generating function for both 
$O(2n)$ and $SO(2n)$, and compute the supersymmetric index for gauge groups of exceptional  
type. A crucial role is played by the classification of nilpotent orbits,
as well as global properties of their centralizers. We give illustrative examples of new features
of our analysis for the $D$-type algebras.

\newpage

{%\small
\tableofcontents }

%\newpage

\section{Introduction}

A basic property of four-dimensional gauge theories that are massive in the infrared is
their number of quantum vacua. This number can for instance be determined for pure 
${\cal N}=1$ supersymmetric Yang-Mills theory in four dimensions. Arguments from chiral symmetry breaking correctly predict the count of massive
vacua. The tally is confirmed by calculating
the supersymmetric index \cite{Witten:1982df,Witten:2000nv}.

Note that even for pure ${\cal N}=1$ supersymmetric
Yang-Mills theory, some subtleties remain. In \cite{Witten:2000nv}, the supersymmetric index was computed
by compactification on $T^3 \times S^1$, and an analysis of commuting triples in the gauge group. On the other
hand, it was argued in \cite{Aharony:2013hda} that the number of quantum vacua after compactification on $S^1$ depends
on the global choice of gauge group and the spectrum of line operators. It would therefore be useful to revisit
the analysis of \cite{Witten:2000nv} in light of the new perspective \cite{Aharony:2013hda}.

In this paper, we concentrate on the ${\cal N}=1$ supersymmetric gauge theory in four dimensions
which is a massive deformation of the  ${\cal N}=4$ 
 theory. The three chiral multiplets of the ${\cal N}=4$ theory
are given arbitrary non-zero masses.
We wish
to compute the number of quantum vacua of the resulting ${\cal N}=1^\ast$ theory
semi-classically.  The infrared dynamics of the gauge theory 
will be governed by the infrared dynamics of the
gauge group left unbroken by the vacuum expectation values of the chiral multiplets. A novel feature compared
to pure ${\cal N}=1$ supersymmetric Yang-Mills theory is that there are both massive and massless phases
in the infrared. This renders the calculation of the number of vacua through compactification of space-time
even more
subtle since in ${\cal N}=1^\ast$ theory with generic gauge group $G$, Wilson lines on circles can
for instance lift Coulomb to massive vacua \cite{Bourget:2015cza}. In this paper,
we work directly in $\mathbb{R}^4$, albeit semi-classically.
By carefully classifying 
vacuum expectation values of the three adjoint chiral multiplets, as well as the corresponding unbroken gauge groups and their global properties, we obtain a prediction for the number of massive vacua of ${\cal N}=1^\ast$ on $\mathbb{R}^4$. 

For $A$-type gauge groups, the number of massive vacua was counted in \cite{Donagi:1995cf,Dorey:1999sj},
while for $B,C$ and $D$ type gauge groups, there were some remarks in \cite{Aharony:2000cw}, while 
almost
complete results were presented in \cite{Naculich:2001us}. 
In this paper, we will perform the semi-classical calculation of the supersymmetric index using a different and more efficient technique. It will allow us to complete
the count in the case of the $D$-type gauge groups, and permit us to recuperate laborious mathematical
classification results which enable us to the predict the supersymmetric index for all exceptional gauge groups
as well. These results complete the count of massive vacua of ${\cal N}=1^\ast$ gauge theories 
 in
four dimensions. The main mathematical tool we use are nilpotent orbits.\footnote{
Nilpotent orbits also play a 
role in classifying  surface operators in supersymmetric gauge theories \cite{Gukov:2008sn} and  co-dimension
two defects in $(2,0)$ theories \cite{Chacaltana:2012zy}.}

In section \ref{classificationproblem}, we explain the three-step program for computing the supersymmetric index,
and the role played by nilpotent orbits of the gauge algebra, and their centralizers. In section \ref{classicalcounting}
we recompute the supersymmetric index for $A,B$ and $C$ type gauge groups, and present the calculation for 
gauge groups of type $D$. The exceptional gauge algebras $E,F$ and $G$ are treated in section  \ref{exceptionalcounting}. 
%The role of the center of the gauge group is briefly discussed in section \ref{centerofgroup}. 
Appendix \ref{dualcoxeterapp} has a table of dual Coxeter numbers and
appendix 
\ref{examples} contains examples that illustrate subtleties discussed in generality in the bulk
of the paper.

\section{The Semi-Classical Configurations and the Classification Problem}
\label{classificationproblem}
In this section, we discuss how the classification of vacuum expectation values for the adjoint
scalars in the chiral multiplets of ${\cal N}=1^\ast$ reduces to the problem of the classification
of nilpotent orbits of the complexified Lie algebra $\mathfrak{g}$ of the gauge group.\footnote{
The idea of using nilpotent orbits in the context of classifying vacua in ${\cal N}=1^\ast$ theory was mentioned in \cite{Wyllard:2009ir}. See also \cite{Kac:1999av} for an application of nilpotent orbit theory to a  supersymmetric
index calculation.  }

\subsection{Semi-classical Configurations and $sl(2)$ algebras}

Our starting point is the  ${\cal N}=1^\ast$ super Yang-Mills theory
with compact gauge group $G$
on $\mathbb{R}^4$. The vacua are classified by solving the F-term equations for constant fields, and dividing the 
solution space by the complexified gauge group $G_{\mathbb{C}}$. 
%{From} now on, we will exclusively
%work with the complexified gauge group and will therefore switch notation to $G_{\mathbb{C}}=G$.
The F-term equations arise from
the superpotential which contains the cubic commutator term of ${\cal N}=4$ super Yang-Mills theory as well
as the three quadratic mass terms. The equations dictate that the (rescaled) adjoint scalar fields $X^i$ (where $i \in \{ 1,2,3 \}$)
form an $sl(2)$ algebra:
\begin{eqnarray}
[ X^i , X^j ] &=& {\epsilon^{ij}}_k X^k \, .
\end{eqnarray}
Thus, the scalars provide us with a map from an $sl(2)$ algebra into the complexified Lie algebra $\mathfrak{g}$ of the gauge group. To find the supersymmetric vacua, we are to classify all $sl(2)$ triples inside the Lie algebra $\mathfrak{g}$, up to gauge equivalence. 
Configurations are gauge equivalent if they are mapped to each other  by the adjoint action of the complexified 
gauge group $G_{\mathbb{C}}$ on the Lie algebra $\mathfrak{g}$. 
%This classification will provide us with
%inequivalent semi-classical configurations. 
Thus, our first step is to review what is known about the classification of 
inequivalent $sl(2)$ triples embedded in the adjoint representation.
%We will come to
%the second step in section \ref{centralizers}.

\subsection{The Gauge Group, Triples and Nilpotent Orbits}

{From} now on, we will denote the complexified gauge group as $ G_{\mathbb{C}} \equiv G$. We need to
make a distinction between various groups that have the same Lie algebra.
One canonical group associated to the Lie algebra $\mathfrak{g}$ is the adjoint group $G_{ad} = \mathrm{Aut}(\mathfrak{g})^o$, namely the identity component of the group of automorphisms of the Lie algebra $\mathfrak{g}$. The adjoint group $G_{ad}$ is alternatively characterized by the fact that it is the group
with algebra $\mathfrak{g}$ and trivial center.

We can now lay the groundwork for the first classification problem. Note that amongst our complex adjoint fields $X^i$, we can identify a linear combination
$X^+$ which is nilpotent by the equations of motion. Indeed, we can consider the complex
combinations $X^{\pm} = \pm X^1 + i X^2$ and $X^0 = 2 i X^3$. Then the non-vanishing commutation relations amongst
these fields are
\begin{equation}
\begin{array}{rcl}
\left[X^0 , X^+\right] &=& \, \,  2 X^+ \\
\left[X^0 , X^-\right] &=& - 2 X^- \\
\left[X^+ , X^-\right] &=& \, \, X^0 . 
\end{array}
\end{equation}
Of course, these remain standard commutation relations of the algebra $sl(2)$. We can describe them by
stating that $(X^0 , X^+ , X^-)$ form an $sl(2)$ triple. The vacuum expectation value $X^+$ is now a nilpotent element of the Lie algebra $\mathfrak{g}$ of the gauge group. Reciprocally, given a nilpotent element in a complex semisimple Lie algebra $\mathfrak{g}$, the Jacobson-Morozov theorem states that we can always extend it to an $sl(2)$ triple. The relation between nilpotent elements and $sl(2)$ triples is  a bijection in the following sense : there is a one-to-one correspondence between $G$-conjugacy classes of $sl(2)$ triples in $\mathfrak{g}$ and non-zero nilpotent $G$-orbits in $\mathfrak{g}$. This follows for instance from Theorem 3.2.10 in \cite{CM} when $G=G_{ad}$, and it remains true for connected gauge groups of any isogeny type (i.e. with non-zero center) because the adjoint action of the center is trivial.
Moreover, if $G$ and $G'$ are two connected groups with the same Lie algebra, then $G$-conjugacy classes and $G'$-conjugacy classes of $sl(2)$ triples in this Lie algebra coincide, as do their nilpotent orbits. For instance, the $SO(2n)$ and $PSO(2n)$ classes of triples and orbits are the same. The assumption that the groups are connected is essential, as we will see in detail in the case of $O(2n)$ and $SO(2n)$. 

The bottom line is that it will be sufficient for us to study
%the centralizers of the $sl(2)$ triples associated to 
the nilpotent orbits of $\mathfrak{g}$ in order to enumerate gauge inequivalent vacuum configurations for the triplet of adjoint scalars in the chiral multiplets. These nilpotent orbits are finite in number and they have been classified \cite{CM,C,LS,LT}. Thus, we have virtually completed the first step.

\subsection{The Centralizer and the Index}

The second step in our program is to determine the unbroken gauge group in a given semi-classical configuration for the adjoint scalar fields. Thus, for each nilpotent orbit with its associated $sl(2)$ triple, we need to determine the centralizer of the triple, i.e. the unbroken gauge group. This determination has been performed as well \cite{CM,C,LS,LT}. The results
are lined with intricacies that we will discuss in due course. Before we do so, we introduce some notation for the relevant
mathematical objects. 
We denote by $C_G(n)$ the centralizer in $G$ of a nilpotent element $n\in \mathfrak{g}$. We call $\mathfrak{trip}$ the image in $\mathfrak{g}$ of a $sl(2)$ triple associated to the $G$-orbit of the nilpotent element $n$. We can then define $C_{G}(\mathfrak{trip})$, the centralizer of $\mathfrak{trip}$ in the group $G$. We are primarily interested in the structure of $C_{G}(\mathfrak{trip})$, the unbroken gauge group in a given semi-classical configuration for the adjoint scalar fields. It will be important to us that in general this group is not connected, and that therefore its component group $Comp(\mathfrak{trip})=C_G(\mathfrak{trip})/C_G(\mathfrak{trip})^o$ is non-trivial. It is crucial for the computation of the supersymmetric index to understand both the structure of the part of the centralizer connected to the identity \emph{and} the action of the component group on the connected components. 

The third step will be to apply our knowledge of the infrared dynamics of pure ${\cal N}=1$ supersymmetric gauge theories to the theory with gauge group $C_{G}(\mathfrak{trip})$. If the unbroken gauge group 
$C_{G}(\mathfrak{trip})$ is non-abelian,
the number of massive vacua on $\mathbb{R}^4$ will be given by the product of the dual Coxeter numbers of the simple factors.\footnote{The number of vacua for pure ${\cal N}=1$ super Yang-Mills theory on $\mathbb{R}^4$ is independent of the center of the gauge group.
%This can for instance be viewed as a consequence of the fact that the path integral is well-defined and that
%all fields transform in the adjoint.
} The set of vacua possibly still needs to be divided out by the action of the component group $Comp(\mathfrak{trip})$.

\section{The Counting for the Classical Groups}
\label{classicalcounting}
In this section, we apply the three-step program set out in section \ref{classificationproblem} to the case
of all classical gauge groups of types $A,B,C$ and $D$. 
\subsection{The Nilpotent Orbits}
In a first step, we describe the nilpotent orbits of the classical Lie algebras $A_{n-1}=sl(n)$,
$B_n=so(2n+1)$, $C_n=sp(2n)$ and $D_n=so(2n)$. 
\begin{table}
\centering
\begin{tabular}{|c|c|}
\hline
Lie Algebra & Labelling of Nilpotent Orbits \\
\hline
$sl(n)$ & Partitions of $n$ \\
\hline
$so(2n+1)$ & Partitions of $2n+1$ in which even parts occur with even multiplicity \\
\hline
$sp(2n)$ & Partitions of $2n$ in which odd parts occur with even multiplicity \\
\hline
$so(2n)$ & Partitions of $2n$ in which even parts occur with even multiplicity \\
 & with \emph{very even} partitions corresponding to two orbits \\
 \hline
\end{tabular}
\caption{The nilpotent orbits in the Lie algebra $\mathfrak{g}$ in the left column are in one-to-one correspondence with the set of partitions described in the right column. The  orbits are assumed to be generated by a 
connected group with algebra $\mathfrak{g}$.}
\label{Partitions}
\end{table}
The nilpotent orbits can be represented in a variety of ways. For the classical algebras,
the nilpotent orbits in connected groups are in one-to-one correspondence with sets of partitions given explicitly in table \ref{Partitions}. Since partitions will play an important role, we introduce a few useful notations. For any partition of an integer $N = d_1 + d_2 + ... + d_N$, we define $n_d = |\{k | d_k = d\}|$, namely the number of times the
integer $d>0$ appears in the partition, as well as the sets 
 \begin{equation}
 D_o = \{ d | d \textrm{ odd and } n_d > 0\}
 \end{equation}
  \begin{equation}
 D_e = \{ d | d \textrm{ even and } n_d > 0\} \, .
 \end{equation}
The sets $D_o$ (respectivale $D_e$) are the sets of odd (respectively even) integers that appear in the partition of $N$. 

\begin{figure}
\centering
\includegraphics[width=0.4\textwidth]{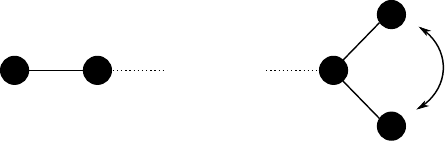}
\caption{The action of the $\mathbb{Z}_2$ outer automorphism of the $D_n$ algebra (for $n>4$) on the Dynkin
diagram.}
\label{OuterAutD}
\end{figure}

As can be seen in table \ref{Partitions}, there is a subtlety for $so(2n)$. We say that a given partition of an integer is \emph{very even} if it has only even parts, each occurring with even multiplicity. A very even partition corresponds to precisely two inequivalent orbits of $so(2n)$. These orbits are interchanged under the outer $\mathbb{Z}_2$ automorphism that acts by exchanging the two extremal nodes on the fork of the $D$-type Dynkin diagram, 
as illustrated in figure \ref{OuterAutD}. This subtlety disappears if instead of $SO(2n)$ orbits, one considers $O(2n)$ orbits : in this case, {\em every} partition of $2n$ with even parts occurring with even multiplicity correspond to a single orbit.\footnote{We note that the classification of orbits in \cite{Naculich:2001us}
for $D$-type gauge groups therefore corresponds to the choice of gauge group $G=O(2n)$.} It will be useful 
to include the gauge group $O(n)$ into our discussion in the following, since it serves as a stepping stone to obtain results
for $SO(n)$ gauge groups.

\subsection{The Centralizers}
\label{centralizers}

For each partition, the centralizer $C_{G}(\mathfrak{trip})$ of the corresponding triple 
is explicitly known \cite{CM,C,LS,LT}. We will make particular choices for the center of the
gauge group, which influence the centralizers, but the final result will be independent of this choice.
Since the massive vacua of the $A$-type ${\cal N}=1^\ast$ 
theory are well-understood \cite{Donagi:1995cf,Dorey:1999sj}, we concentrate on type $B,C$ and $D$ gauge algebras. To describe the centralizers 
of the triple for the gauge groups $(S)O(n)$ and $Sp(2n)$ we introduce more notation. Firstly, if  $H$ 
is a  matrix group, we use the notation $H^p_{\Delta}$ to denote the diagonal copy of the group $H$ inside the direct product group $H^p$. Secondly, we denote by $S\left( \prod\limits_{d=1}^m H_d\right)$ the subgroup of the product of matrix groups $\prod\limits_{d=1}^m H_d$ consisting of $m$-tuples of matrices whose determinants have product one.

\subsubsection{The Centralizer in $O(n)$ and $SO(n)$}
The basic result for the centralizer of a triple associated to a nilpotent orbit (labelled by a partition satisfying
$\sum_{d} n_d \,  d = n$)
in $O(n)$ can then be written as
\begin{equation}
C_{O(n)}(\mathfrak{trip}) = \prod\limits_{d \in D_e} \left[ Sp(n_d) \right]^d_{\Delta} \times \prod\limits_{d \in D_o} \left[ O(n_d) \right]^d_{\Delta} , 
\end{equation}
which is isomorphic to the group
\begin{equation}
\label{centralizerOn}
C_{O(n)}(\mathfrak{trip})  \cong \prod\limits_{d \in D_e}  Sp(n_d)  \times \prod\limits_{d \in D_o} O(n_d) . 
\end{equation}
The second expression (\ref{centralizerOn}) is simpler, but the first one realises the group $C_{O(n)}(\mathfrak{trip})$ as an explicit subgroup of $O(n)$ in the fundamental representation. 
{From} the first expression then,   it is 
straightforward to derive the centralizer in $SO(n)$. Indeed we merely have to enforce the constraint that the determinant be 1, which gives: 
\begin{equation}
C_{SO(n)}(\mathfrak{trip}) = S \left( \prod\limits_{d \in D_e} \left[ Sp(n_d) \right]^d_{\Delta} \times \prod\limits_{d \in D_o} \left[ O(n_d) \right]^d_{\Delta}   \right) \, .
\end{equation}
We note that in a generic situation, the centralizer group is not connected, even if we start out with a connected
gauge group.
For further clarity, we construct an explicit matrix representation of the centralizer subgroup. For each dimension $d$ such that $n_d >0$, take a matrix $M_{n_d}$ in $Sp(n_d)$ if the dimension $d$ is even and in $O(n_d)$ if $d$ is odd. Then build 
the matrix $M = Diag(M_{n_d}^d)$ (i.e. where the matrix $M_{n_d}$ is present $d$ times along the diagonal). 
The centralizer is then the group of such matrices $M$ with determinant 1. It then is manifest that
\begin{equation}
\label{centralizerSOn}
C_{SO(n)}(\mathfrak{trip}) \cong \prod\limits_{d \in D_e}  Sp(n_d)  \times S \left(  \prod\limits_{d \in D_o}  O(n_d) \right) \, .
\end{equation}
Let us show more explicitly that the centralizer is not necessarily connected. For any matrix $M$ constructed as above, 
\begin{equation}
\det M = \prod\limits_{d \in D_e \cup D_o} (\det M_{n_d})^d = \prod\limits_{d \in D_o} \det M_{n_d}
\end{equation}
If the set of odd dimensions appearing in the partitions is not empty, $D_o \neq \emptyset$, the constraint on the determinant removes half of the connected components, leaving $2^{|D_o|-1} %\geq 1
$ connected components. When $D_o = \emptyset$, the constraint is automatically enforced. Hence we have
\begin{equation}
\label{componentSOn}
Comp_{SO(n)}(\mathfrak{trip}) = \mathbb{Z}_2^{\mathrm{max} (0,|D_o|-1)} . 
\end{equation}

\subsubsection{The centralizer in $Sp(2n)$}
Similarly, the centralizer of a triple inside $Sp(2n)$ is given explicitly by
\begin{equation}
\label{centralizerSp2n}
C_{Sp(2n)}(\mathfrak{trip}) =  \prod\limits_{d \in D_o} \left[ Sp(n_d) \right]^d_{\Delta} \times \prod\limits_{d \in D_e} \left[ O(n_d) \right]^d_{\Delta} 
\end{equation}
and is isomorphic to 
\begin{equation}
C_{Sp(2n)}(\mathfrak{trip}) \cong \prod\limits_{d \in D_o} Sp(n_d) \times \prod\limits_{d \in D_e} O(n_d) . 
\end{equation}
The component group is equal to:
\begin{equation}
\label{componentSp2n}
Comp_{Sp(2n)}(\mathfrak{trip}) =  \mathbb{Z}_2^{|D_e|} . 
\end{equation}

\subsection{The Supersymmetric Index for the Classical Groups}
\label{indexcalculationclassicalgroups}
We have determined the set of inequivalent semi-classical configurations for the adjoint scalar fields $X^i$,
as well as  the subgroup of the gauge group that is left unbroken by the vacuum expectation values. To compute the semi-classical number of massive vacua, we compute the Witten indices of the pure ${\cal N}=1$ theories that arise upon fixing a given semi-classical configuration for the fields $X^i$, and add them up for all possible inequivalent semi-classical configurations. The global properties of the unbroken gauge group come into play at this stage -- 
we have already stressed that generically, the unbroken gauge group will not be connected (even if we started out
with a connected gauge group).
We start with a demonstration of how to take into account this complication in the elementary case of the groups $O(4)$ and $SO(4)$. 
%Then we move on to the other connected groups and finally give an explicit formula for the Witten index of pure %${\cal N}=1$ with gauge group the centralizer of any nilpotent element in any classical group. 
With this example in mind, we can generalize this third step and compute the supersymmetric index for all 
${\cal N}=1^\ast$ theories with classical gauge groups.

\subsubsection{$O(4)$ and $SO(4)$}

We firstly recall that we have the equivalence of groups $SO(4)=(SU(2) \times SU(2))/\mathbb{Z}_2$ where
the element we divide out by is the diagonal center $(-1,-1)$ in the product of the $SU(2)$ groups. 
%Translated into the gauge theory language, this means simply that going from gauge group $SO(4)$ to gauge group %$O(4)$ amounts to identifying the two $SU(2)$ factors.  
Secondly, the $O(4)$ group contains two components, and in particular, it contains a parity operation that
exchanges the two $su(2)$ algebras that make up the $so(4)$ algebra. This can be viewed as a 
special case of the  $\mathbb{Z}_2$ outer automorphism operation on the $so(2n)$ Dynkin diagram
of figure \ref{OuterAutD}.

The extra gauge symmetry present in the $O(4)$ theory has direct consequences for the counting of inequivalent
vacua.
For a pure ${\cal N}=1$ supersymmetric gauge theory with gauge group $SU(2)$, the number of ground states is two, and they are labelled by a gaugino bilinear  $\langle \lambda \lambda \rangle = \pm 1$. When the gauge group is $SU(2) \times SU(2)$, we therefore have $2 \times 2 = 4$ ground states. %$\langle \lambda \lambda \rangle = (\langle \lambda_1 \lambda_1 \rangle , \langle \lambda_2 \lambda_2 \rangle) = (\pm 1 \pm 1)$. 
In pure ${\cal N}=1$ with gauge group $SO(4)$, the number of ground states remains $4$, because the action of the diagonal center $\mathbb{Z}_2$ is trivial on each fermion bilinear.
%$\langle \lambda_j \lambda_j \rangle$ ($j=1,2$). 
The theory with gauge group $O(4)$, on the other hand has the extra element in the gauge group that  exchanges the two $SU(2)$ gauge group factors. We must restrict to those vacua that are invariant under this element of the gauge group as well, and these are schematically represented as $(+,+)$, $(-,-)$ and $\frac{1}{\sqrt{2}} \left( (+,+) + (-,-) \right)$. Thus, the Witten index is three for the $O(4)$ theory. The component group matters.
%(The same would be true for a $Spin(4)$ versus a $Pin(4)$ gauge group.)

\subsubsection{The Consequences of the Component Group}
In fact, the example of $O(4)$ is central for the following reason. 
We know that for a pure ${\cal N}=1$ theory with connected simple Lie group as a gauge group, 
the Witten index is given by the dual Coxeter number of the underlying Lie algebra. (See appendix 
\ref{dualcoxeterapp} for the relevant table.) If the gauge group is not connected, but has a gauge algebra with a connected
Dynkin diagram, then the Witten index remains unchanged. (The extra elements of the gauge group leave the fermion bilinear invariant.) Thus, the only non-trivial case that we need to keep in mind when studying the 
%
%Since a connected semisimple Lie group is simple if and only if its Dynkin diagram is connected, this gives the %answer for all gauge groups $Sp(2n)$ and also for all $SO(n)$, with the exception of $SO(4)$ whose Dynkin diagram is %disconnected. Note that these are the only groups that appear in the %
%
centralizers (\ref{centralizerSOn}) and (\ref{centralizerSp2n}) is precisely the case of $O(4)$. All other
cases correspond to either trivial groups, a Coulomb vacuum (if an $SO(2)$ factor is present), or the product
of simple factors for whom the component group will not influence the Witten index. Thus,
we must keep an eye out only for the difference between the disconnected Dynkin diagram of 
$so(4)=su(2) \oplus su(2)$ which can be subject or not to an exchange identification, depending on the component group of the centralizer.

\subsubsection{The Contribution of Each Centralizer}

We apply the results derived for the simple groups to the centralizers (\ref{centralizerSOn}) and (\ref{centralizerSp2n}). The general strategy is first to compute the number of vacua in pure $\mathcal{N}=1$ with gauge group equal to the connected component of the centralizer, and then modify this counting if necessary 
by taking into account the action of the component groups  (\ref{componentSOn}) and (\ref{componentSp2n}). 

 As a crucial warm-up exercise,
 we consider the case of the gauge group
 $S ( O(4)^m )$ for $m \geq 1$. We wish to compute the number of vacua for a pure ${\cal N}=1$ theory with
 this gauge group. The gauge group has $2^{m-1}$ connected components, all isomorphic 
 (as manifolds) to $SO(4)^m$. We can write 
 \begin{equation}
 S \left( O(4)^m \right) = SO(4)^m \rtimes \mathbb{Z}_2^{m-1} = \left( \frac{SU(2) \times SU(2)}{\mathbb{Z}_2}\right)^m \rtimes \mathbb{Z}_2^{m-1}  \, .
 \end{equation}
 The Witten index for gauge group $SO(4)^m$ is $4^m$. Now we have to take into account the $\mathbb{Z}_2^{m-1}$ factor. This group
 acts by exchanging an even number of %pairs of 
 $SU(2)$ factors in the product above. Then the gauge invariant
 vacua, described in terms of $SU(2)$ factor vacua, are enumerated as follows : 
 \begin{itemize}
 \item 2 configurations in which all summands consist of factors equal to $(+-)$ or $(-+)$ (namely, $(+-)^m$ and $(-+)(+-)^{m-1}$, symmetrized appropriately)
 \item $3^m-1$ configurations containing at least one $(++)$ or one $(--)$ (namely, all words with $m$ letters chosen from the alphabet $\{ (++),(+-),(--) \}$, except the word $(+-)^m$).  
 \end{itemize}
 In total, this gives  $3^m+1$ gauge invariant vacua for gauge group $S(O(4)^m)$.

We have now gathered all elementary ingredients to proceed with the computation of the contribution to the Witten index of each given nilpotent orbit with associated centralizer. Let's start with the centralizer (\ref{centralizerSp2n}) for gauge group $Sp(2n)$. The identity component is isomorphic to 
\begin{equation}
\prod\limits_{d \in D_o} Sp(n_d) \times \prod\limits_{d \in D_e} SO(n_d)
\end{equation}
and the component group (\ref{componentSp2n}) ignores the $Sp(n_d)$ factors and acts on all the $SO(n_d)$ factors. In particular, the $SU(2)$ components inside the $SO(4)$ factors are exchanged independently, and the corresponding index is always $3$. Taking into account all the factors, we conclude that the Witten index is 
\begin{equation}
\prod\limits_{d \in D_o} \left( \frac{n_d}{2} +1 \right) \prod\limits_{d \in D_e} I(n_d)
\end{equation}
where the function $I$ counts the number of massive vacua for a pure ${\cal N}=1$ theory with
gauge group $O(n)$.

Next, we turn to the centralizer (\ref{centralizerSOn}) for the $SO(n)$ nilpotent orbits.
The identity component is
\begin{equation}
\label{centralizerSO2np1connected}
\prod\limits_{d \in D_e}  Sp(n_d)   \prod\limits_{d \in D_o}  SO(n_d) 
\end{equation}
and the component group (\ref{componentSOn}) acts on the $O$ factors in the centralizer  (\ref{centralizerSOn})
as follows :  we can change the sign of the determinant of an even number of matrices in the $O(n_d)$ factors. If there are no $O(4)$ factors, the index follows immediately.
%from (\ref{WittenIndexSimpleGroups}). 
The opposite extreme, in which all factors of type $O(n_d)$ are $O(4)$ factors (and there is at least one such $O(4)$ factor), we have treated in our warm-up exercise, and we just take the contribution of the $Sp(n_d)$ factors into account to obtain in that case
\begin{equation}
\left( 3^{|D_o|}+1 \right)   \prod\limits_{d \in D_e} \left( \frac{n_d}{2} +1 \right) . 
\end{equation}
Note that in this case, $N = \sum\limits_{d \in D_e}  n_d \, d + 4 \sum\limits_{d \in D_o} d $ is even
by construction.
Thus, this particular case arises only for algebras of type $D$ i.e. $SO(2n)$ gauge groups. 

Finally, if there exists at least one $n_d \in D_o$ such that $n_d \neq 4$, then we can change the sign of the determinant of any number (odd or even) of matrices in the $O(4)$ factors. Hence the index is again given
by the naive formula
\begin{equation}
 \prod\limits_{d \in D_e} \left( \frac{n_d}{2} +1 \right) \prod\limits_{d \in D_o} I(n_d) \, .
\end{equation}
The supersymmetric indices for the centralizers for all classical groups are summarised 
in table \ref{ClassicalWittenIndices}.
{\renewcommand{\arraystretch}{1.2}
\begin{table}
\begin{tabular}{|c|c|c|c|}
\hline
Type & Unbroken Gauge Group & {Case} & {Index}  \\
\hline
$B,D$ & 
{\small 
 $S \Big( \prod\limits_{d \in D_e} \left[ Sp(n_d) \right]^d_{\Delta} \times \prod\limits_{d \in D_o} \left[ O(n_d) \right]^d_{\Delta}   \Big) $ }
 & 
%$ \begin{array}{c}
%   |D_o \cap N_4|=0 \textrm{ or }   
%\\
% \left( |D_o \cap N_4| \neq 0 \text{ and } |D_o \cap \bar{N}_4| \neq 0 \right) 
% \end{array}
% $
\begin{tabular}{c}
No $O(4)$ factor \\
or 
\\
at least one $O(n)$ factor 
\\
with  $n \ge 1$ and $n \neq 4$.
% Let's discuss the case n=1 on occasion.
\end{tabular}
 &
 $
 \prod\limits_{d \in D_e} \left( \frac{n_d}{2} +1 \right) \prod\limits_{d \in D_o} I(n_d) 
 $ 
 \\
 \hline
$ D$ 
& 
$ 
S \big( \prod\limits_{d \in D_e} \left[ Sp(n_d) \right]^d_{\Delta} \times \prod\limits_{d \in D_o} \left[ O(4) \right]^d_{\Delta}   \big) 
$ 
&
%
%$
%|D_o \cap N_4| \neq 0 \text{ and } |D_o \cap \bar{N}_4| = 0 
%$
Only $O(4)$ factors.
&
$
\left( 3^{|D_o|}+1 \right)   \prod\limits_{d \in D_e} \left( \frac{n_d}{2} +1 \right)  
$  
\\
 \hline
$C$ & $ \prod\limits_{d \in D_o} \left[ Sp(n_d) \right]^d_{\Delta} \times \prod\limits_{d \in D_e} \left[ O(n_d) \right]^d_{\Delta} 
$ &
{No restriction}  &
$\prod\limits_{d \in D_o} \left( \frac{n_d}{2} +1 \right) \prod\limits_{d \in D_e} I(n_d)  $
\\
\hline
\end{tabular}
\caption{The supersymmetric index contributions for unbroken subgroups in the connected gauge groups 
$SO(2n+1)$, $Sp(2n)$ and $SO(2n)$. The function $I$ counts the number of massive vacua for a pure ${\cal N}=1$ theory with
gauge group $O(n)$, and is given explicitly by $I(1)=1$, $I(3)=2$, $I(4)=3$ and $I(n)=n-2$ for $n \neq 1,3,4$. }
\label{ClassicalWittenIndices}
\end{table}

 \subsection{The Generating Functions}
 \label{genfunct}
 In this subsection, we write down the generating function for the supersymmetric indices that
 we have computed in subsection \ref{indexcalculationclassicalgroups}.
For $A_{n-1}$-type Lie algebras, no subtleties arise and the supersymmetric index is the sum of the divisors of $n$ \cite{Donagi:1995cf,Dorey:1999sj}. A generating function is therefore $I_{SU(n)}=\sum_{n=1}^\infty \sigma_1(n) q^n$. 

Next, 
we write down the $SO(n)$ generating function. Firstly, we refer to 
a detailed discussion of the generating function for partitions satisfying the 
constraint that even parts occur with even multiplicity in \cite{Wyllard:2007qw}, and we have derived that
it codes the Witten index contribution
for the centralizer in $O(n)$. We therefore state that the generating function of
\cite{Wyllard:2007qw} captures the number 
of vacua in the ${\cal N}=1^\ast$ theory with $O(n)$ gauge group. 
The supersymmetric index of the $O(n)$ theory
is  then the coefficient of $q^n$ in the series expansion of 
\begin{equation}
\label{GenFuncOn}
I_{O(n)}(q) = \prod\limits_{k=1}^{\infty} \frac{P_0(q^{2k-1})}{(1-q^{2k-1})^2 (1-q^{4k})^2} 
\end{equation}
where $P_0(x) = 1-x-x^2+3x^3-x^4-x^5+x^6$.

To write down the generating function for gauge group $SO(n)$, we will need to take into account the doubling of the very even partitions and the fact that the centralizers in $SO(n)$ satisfy the constraint that their overall determinant is equal to one. The very even partitions are made out of elementary blocks of the form $2k + 2k + ... + 2k$ with $2m$ terms, and they contribute a factor $Sp(2m)$ in the residual symmetry, giving rise to $m+1$ 
quantum vacua. The corresponding contribution in the partition of $n$ is $4mk$, so the number of vacua corresponding to the very even partitions is the coefficient of $q^n$ in the generating function 
\begin{equation}
\label{GenFuncVE}
\prod\limits_{k=1}^{\infty} \sum\limits_{m=0}^{\infty} (m+1) q^{4km} = \prod\limits_{k=1}^{\infty} \frac{1}{(1-q^{4k})^2} %= 1+2q^4 + 5q^8 + 10 q^{12} + ... 
\, .
\end{equation}
A partition is either very even, or not. Moreover, the very even partitions are already counted once in the generating
function (\ref{GenFuncOn}) -- it is the origin of the second factor in the denominator \cite{Wyllard:2007qw}.
Thus, to count them with the required double multiplicity, we add the generating function (\ref{GenFuncVE})
to the generating function (\ref{GenFuncOn}), obtaining
\begin{equation}
\label{GenFuncOn+VE}
\prod\limits_{k=1}^{\infty} \frac{1}{(1-q^{4k})^2} \left( 1+\prod\limits_{k=1}^{\infty} \frac{P_0(q^{2k-1})}{(1-q^{2k-1})^2} \right) \,  . 
%= 2+q+3 q^3+8 q^4+6 q^5+7 q^6+15 q^7%+31 q^8 
\end{equation}
However, we still have to take special care of the partitions that fall under the restrictions of 
the second line in table \ref{ClassicalWittenIndices}. In the expression (\ref{GenFuncOn+VE}), all such partitions received an index $3^{|D_o|}  \prod\limits_{d \in D_e} \left( \frac{n_d}{2} +1 \right) $ (corresponding to imposing all $O(4)$ gauge invariances), and we must add those vacua that are gauge invariant under the smaller unbroken gauge group arising 
from the unit determinant requirement. To add this contribution, we first write the generating function of the number of partitions that give rise to an unbroken $S \left( O(4)^m\right)$ gauge group factor, possibly
supplemented with $Sp(2n)$ gauge group factors. These are partitions of $N=2n$ into odd integers, each of which appear exactly four times. Equivalently, they are partitions of $N/4$ into distinct odd integers. 
The coefficient of $x^n$ in 
\begin{equation}
-1 + \prod\limits_{k=1}^{\infty} (1+x^{2k-1})
\end{equation}
is the number of partitions of $n$ into distinct odd integers, so that setting $x=q^4$ gives us the generating
function for this partition problem. Including the supplementary $sp(2n)$ factors
%$\prod\limits_{d \in D_e} \left( \frac{n_d}{2} +1 \right) $ factor 
boils down to multiplying by the generating function (\ref{GenFuncVE}), and we thus get the extra
gauge invariant massive vacua counting function: 
\begin{equation}
\label{GenFuncS}
\prod\limits_{k=1}^{\infty} \frac{1}{(1-q^{4k})^2} \left(-1+ \prod\limits_{k=1}^{\infty} (1+q^{8k-4}) \right) \, .
\end{equation}
Finally, the generating function for $SO(n)$ follows from adding the $O(n)$ generating function,
supplemented with the second copy of the very even partitions coded in (\ref{GenFuncOn+VE}) as well as the
extra gauge invariant vacua of equation (\ref{GenFuncS}). We obtain that the number of massive vacua in the $SO(n)$ theory is the coefficient of $q^n$ in 
\begin{equation}
\label{GenFuncSOn}
I_{SO(n)} (q) = \prod\limits_{k=1}^{\infty} \frac{P_0(q^{2k-1})}{(1-q^{4k})^2(1-q^{2k-1})^2} + \prod\limits_{k=1}^{\infty} \frac{1+q^{8k-4}}{(1-q^{4k})^2} \, .
\end{equation}
The generating function for $SO(2n+1)$  gauge group is obtained by taking the odd part
\begin{equation}
\label{GenFuncO2np1}
I_{SO(2n+1)}(q) = I_{O(2n+1)}(q) =   \frac{1}{2} \prod\limits_{k=1}^{\infty} \frac{P_0(q^{2k-1})}{(1-q^{2k-1})^2(1-q^{4k})^2} - \frac{1}{2} \prod\limits_{k=1}^{\infty} \frac{P_0(-q^{2k-1})}{(1+q^{2k-1})^2(1-q^{4k})^2}   \, . 
\end{equation}
This result was obtained in \cite{Wyllard:2007qw}. In the case of $B$-type gauge group, the counting function is 
identical for $O(2n+1)$ and $SO(2n+1)$.
For the $D$-type gauge groups, we find
\begin{equation}
\label{GenFuncO2n}
I_{O(2n)}(q) =   \frac{1}{2} \prod\limits_{k=1}^{\infty} \frac{P_0(q^{2k-1})}{(1-q^{2k-1})^2(1-q^{4k})^2} 
+ \frac{1}{2}\prod\limits_{k=1}^{\infty} \frac{P_0(-q^{2k-1})}{(1+q^{2k-1})^2(1-q^{4k})^2} \, ,
\end{equation}
and 
\begin{equation}
I_{SO(2n)} (q) = \frac{1}{2}\prod\limits_{k=1}^{\infty} \frac{P_0(q^{2k-1})}{(1-q^{4k})^2(1-q^{2k-1})^2} + \frac{1}{2}\prod\limits_{k=1}^{\infty} \frac{P_0(-q^{2k-1})}{(1-q^{4k})^2(1+q^{2k-1})^2} + \prod\limits_{k=1}^{\infty} \frac{1+q^{8k-4}}{(1-q^{4k})^2}  \, . 
\end{equation}
The last term in equation (\ref{GenFuncSOn}) contributes to the $SO(2n)$ generating function only.
In appendix \ref{Dexamples} we discuss a few detailed examples that illustrate the special features of the $D_{n}$
supersymmetric index.

Finally, for reference we also give the generating function for $Sp(2n)$: 
\begin{equation}
I_{Sp(2n)} (q) = \prod\limits_{k=1}^{\infty} \frac{P_0 (q^{2k})}{(1-q^{2k})^2 (1-q^{4k-2})^2} = I_{SO(2n+1)}(q)
\, .
\end{equation}
The last equality is non-trivial and was proven in \cite{Wyllard:2007qw}. It is a consequence of the S-duality
between $B$ and $C$-type ${\cal N}=1^\ast$ gauge theories. The identity of the left and right hand side
was already
noted by Ramanujan.

\section{The Counting for the Exceptional Groups}
\label{SectionExceptional}
\label{exceptionalcounting}
In this section, we count the number of massive vacua for mass-deformed ${\cal N}=4$ super Yang-Mills theory
with exceptional gauge group. For determining the centralizer subgroups, we assume that our gauge group is the adjoint
group $G=G_{ad}$. The supersymmetric indices for other choices of centers are identical.

\subsection{The Orbits and the Centralizers}
The first two steps in our program consist of listing the nilpotent orbits, in bijection with the $sl(2)$ triples,
and the centralizers, the subgroup of the gauge group left unbroken by the adjoint scalar field vacuum expectation
values. While there is a handy list of nilpotent orbits (see e.g. \cite{C}) of exceptional Lie algebras available, 
we need to delve slightly deeper
into the mathematics to understand the centralizers of the associated triples.

Let $n$ be a nilpotent element of a Lie algebra $\mathfrak{g}$ and $\mathfrak{trip}$ the span of an $sl(2)$ triple
corresponding to $n$. The centralizer of the triple is reductive (i.e. semi-simple plus abelian factors) and is a factor in the centralizer of the nilpotent element $C_G(n)=C_G(\mathfrak{trip}) \ltimes U$ where $U$ is the unipotent radical of $C_G(n)$. 
%We have that the unipotent radical is connected and that the quotient groups of the whole group divided by the connected part is the same for $C_G(n)$ and $C_G(\mathfrak{trip})$ as a consequence.
From chapter 13 of \cite{C} we can read off both the type and the component group of $C_G(\mathfrak{trip})$ (called $C$ there), which is almost all we need to compute the contribution to the Witten index for a given orbit. However, we still have to know precisely how the component group of $C_G(\mathfrak{trip})$ acts on its connected components. These are final gauge equivalence identifications that we will need to perform on the vacua of effective pure ${\cal N}=1$ supersymmetric Yang-Mills theories. These  actions can be deduced from the detailed reference \cite{LT}. 

In the following subsection, %\ref{sectionIndexExceptional}, 
we explicitly compute the supersymmetric index by going through the lists of nilpotent orbits, following the order of the lists provided in \cite{C},\footnote{The numbering used in \cite{LT} is one lower than our numbering.} and for each of these orbits, we compute the contribution to the index as follows. 
%If the orbit falls into one of the three configurations (a),(b) or (c) listed below, it is readily computed: 
\begin{itemize}
\item [(a)] 
If the unbroken gauge group $C_G(\mathfrak{trip})$ contains an abelian factor, we have a massless,
Coulomb vacuum, and the contribution is zero. 
\item
[(b)] 
If the unbroken gauge group $C_G(\mathfrak{trip})$ is simple (respectively trivial), the contribution is the dual Coxeter number of the corresponding Lie algebra (respectively one).  (The action of a possible component group on the 
gaugino bilinear is trivial.)
\item
[(c)] 
If the unbroken gauge group $C_G(\mathfrak{trip})$ contains several simple factors, and the component group is trivial, then  the contribution is the product of the dual Coxeter numbers of the simple factors. 
\item 
[(d)]
If the unbroken gauge group $C_G(\mathfrak{trip})$ contains several simple factors, and the component group is non-trivial, then we consider
the tensor product of the quantum vacua of each simple factor, and count those  vacua which are also gauge invariant with respect
to the action of the component group.
\end{itemize}

\subsection{The Supersymmetric Index for the Exceptional Groups}
\label{sectionIndexExceptional}
We proceed on a case-by-case basis.

\subsubsection*{The algebra $G_2$}
The number of orbits is 5. For each of these orbits, we read off the Lie algebra of the unbroken
gauge group in \cite{C} 
(see our table \ref{G2table}) and observe that they are all simple. Hence the supersymmetric index is:
\begin{equation}
I_{G_2} = 4+2+2+1+1=10.
\end{equation} 
\begin{table}[H]
\centering
\begin{tabular}{|c|c|c|c|}
\hline
 Number & Unbroken Gauge Algebra & Component group & Number of Massive Vacua
 \\
 \hline
 1 & $G_2$ & 1 & 4 \\
 2 & $A_1$ & 1 & 2\\
 3 & $A_1$ & 1 & 2 \\
 4 & $1$  & $S_3$ & 1 \\
 5 & $1$   & 1 &   1 \\
 \hline
\end{tabular}
\caption{The number of the orbit in the table in \cite{C}, the unbroken gauge algebra,
the non-identity component of the unbroken gauge group and the number of massive vacua the orbit
gives rise to. }
\label{G2table}
\end{table}
\noindent

\subsubsection*{The algebra $F_4$}
There are 16 orbits for the algebra $F_4$, enumerated in table \ref{F4table}. All centralizers are either simple or trivial (case (b) above), 
except for the orbits number 4 and 8.
Note that for orbits number 3 and 5 (as a few examples amongst many), the component group acts non-trivially on the gauge algebra.
The outer automorphism action leaves the fermion bilinear of pure ${\cal N}=1$ super Yang-Mills theory invariant.

\begin{longtable}{|c|c|c|c|c|}
\hline
 Number & Gauge Algebra & Component group  & Action & Massive Vacua \\
 \hline
 \endhead
\hline
\endfoot
 \endlastfoot
 1 & $F_4$ & 1     &  & 9 \\
 2 & $C_3$ & 1     &  & 4\\
 3 & $A_3$ & $S_2$ & 
  \begin{tikzpicture}[baseline=0,scale=.4]
\draw  (0,0) to (2,0);
\draw  (2,0) to (4,0);
\filldraw (0,0) circle [radius=0.08];
\filldraw (2,0) circle [radius=0.08];
\filldraw (4,0) circle [radius=0.08];
\draw[<->] (0.2,0.2) to[bend left] (3.8,0.2);
\end{tikzpicture}
 & 4\\
 4 & $A_1+A_1$ & $1$ & & $2 \times 2$ \\
 5 & $A_2$ & $S_2$ & 
 \begin{tikzpicture}[baseline=0,scale=.4]
\draw  (0,0) to (2,0);
\filldraw (0,0) circle [radius=0.08];
\filldraw (2,0) circle [radius=0.08];
\draw[<->] (0.2,0.2) to[bend left] (1.8,0.2);
\end{tikzpicture}
 & 3\\
 6 & $G_2$ & 1 &  & 4\\
 7 & $A_1$ & 1 & & 2\\
 8 & $A_1 + A_1$ & $S_2$ & 
 \begin{tikzpicture}[baseline=0,scale=.4]
\filldraw (0,0) circle [radius=0.08];
\filldraw (2,0) circle [radius=0.08];
\draw[<->] (0.2,0.2) to[bend left] (1.8,0.2);
\end{tikzpicture}
 & $ 2\times 2 -1$ \\
 9 & $A_1$ & 1 & & 2 \\
 10 & $A_1$ & $S_2$ & & 2\\
 11 & $1$   & $S_4$ & & 1\\
 12 & $A_1$ & 1     & & 2 \\
 13 & $A_1$ & 1     & & 2\\
 14 & $1$   & $S_2$ & & 1\\
 15 & $1$   & $S_2$ & & 1\\
 16 & $1$   & 1 & & 1\\
 \hline
\caption{The number of the orbit in the \cite{C} table, the unbroken gauge algebra,
the non-identity component of the gauge group, as well as its
action on the gauge algebra, in a selection of cases. The last column indicates the contribution to the
supersymmetric index.}
\label{F4table}
\end{longtable}
Orbit number 4 has trivial component group, and therefore falls into case (c) where we 
count all tensor product vacua. More interestingly, in the case of  orbit 8, the Lie algebra of the unbroken gauge group is $A_1 \oplus A_1$, and the disconnected component of the centralizer acts to exchange the two $su(2)$ algebras. This is a familiar phenomenon, and we realize that the number of gauge invariant vacua is $3$. The total number of massive vacua is therefore:
\begin{equation}
I_{F_4} = 9+4+4+(2 \times 2)+3+4+2+(3)+2+2+1+2+2+1+1+1=45.
\label{IF4}
\end{equation}
We have put the contribution of orbit number 4 and orbit number 8 between parentheses in equation
(\ref{IF4}) to clearly exhibit the different
number of vacua which they contribute to the total supersymmetric index, despite the fact that the algebra of the
centralizer is identical for both orbits.

\subsubsection*{The algebra $E_6$}

For the $E_6$ algebra, there are 21 orbits, listed in table \ref{E6table}.
 \begin{longtable}{|c|c|c|c|c|}
\hline
 Number & Gauge Algebra & Component group  & Action & Massive Vacua \\
 \hline
 \endhead
\hline
\endfoot
 \endlastfoot
 1 & $E_6$ & 1 & & 12\\
 2 & $A_5$ & 1 & & 6\\
 3 & $B_3+T_1$ & 1 & & 0 \\
 4 & $A_2+A_1$ & 1 & & $2 \times 3$\\
 5 & $A_2+A_2$ & $S_2$  & 
\begin{tikzpicture}[baseline=0,scale=.4]
 \draw  (0,0) to (2,0);
\filldraw (0,0) circle [radius=0.08];
\filldraw (2,0) circle [radius=0.08];
\draw (0,1) to (2,1);
\filldraw (0,1) circle [radius=0.08];
\filldraw (2,1) circle [radius=0.08];
%\draw[<->] (-0.1,0.2) to[bend left] (-0.1,0.80);
%\draw[<->] (2.1,0.2) to[bend right] (2.1,0.80);
\draw[<->] (1,0.15) to (1,0.85);
\end{tikzpicture}
                  & $3 \times 3 -3$ \\
 6 & $A_2+T_1$  & 1 &  & 0\\
 7 & $G_2$      & 1 & & 4\\
 8 & $A_1+T_1$  & 1 & & 0  \\
 9 & $B_2+T_1$  & 1 & & 0 \\
 10 & $A_1$     & 1 & & 2\\
 11 & $A_1+T_1$ & 1 & & 0 \\
 12 & $T_2$ & $S_3$ & & 0\\
 13 & $A_1+T_1$ & 1 & & 0 \\
 14 & $A_2$     & 1 & & 3\\
 15 & $T_1$     & 1 & &  0\\
 16 & $A_1$     & 1 & & 2\\
 17 & $T_1$     & 1 & & 0\\
 18 & $1$   & $S_2$ & & 1\\
 19 & $T_1$     & 1 & & 0\\
 20 & $1$       & 1 & & 1\\
 21 & $1$       & 1 &  & 1\\
 \hline
\caption{The number of the orbit, the unbroken gauge algebra,
the non-identity component of the gauge group, as well as its
action on the gauge algebra (when relevant), and the resulting number of
massive vacua.}
\label{E6table}
\end{longtable}
\noindent
All orbits have either a simple centralizer, or a trivial component group (and therefore fall
into classes (b) or (c)), except orbit number 5. The component group of orbit number 5 
exchanges the two $su(3)$ algebras in the unbroken gauge group. We therefore need to count
the $SU(3) \times SU(3)$ pure ${\cal N}=1$ super Yang-Mills vacua which are invariant under this
interchange. The  $ 3 \times 3$  representation of the $\mathbb{Z}_2$ exchange splits into
$6$ invariants and $3$ non-trivial representations. We therefore find the index   
\begin{eqnarray}
I_{E_6} &=& 12+ 6+ 0 + 2 \times 3 + (3 \times 3-3) +
0+ 4 + 0 + 0 + 2 \nonumber \\
&&
+ 0+0+0+3+0+2+0+1+0+1+1 
\nonumber \\
&=& 44.
\end{eqnarray}
% Orbit 12 is massless. Orbit 18 is a Higgs vacuum, and therefore there cannot be a non-trivial action of the center. (It might determine the location of the extremum in the quantum theory, but never mind this for now.)

\subsubsection*{The algebra $E_7$}
There are 45 orbits, among which only orbits number 16 and 19 need special care. 
\begin{longtable}{|c|c|c|c|c|}
\hline
 Number & Gauge Algebra & Component group  & Action & Massive Vacua \\
 \hline
 \endhead
\hline
\endfoot
 \endlastfoot
 1 & $E_7$ &  1 & & 18  \\
 2 & $D_6$ &  1 & & 10 \\
 3 & $B_4+A_1$ & 1 & & 14 \\
 4 & $F_4$ &  1 & & 9 \\
 5 & $C_3+A_1$ & 1 &  &  8 \\
 6 & $A_5$ & $S_2$ &  & 6\\
 7 & $C_3$ & 1 & & 4 \\         
 8 & $A_3 + T_1$ & $S_2$ & & 0  \\
 9 & $3 A_1$ & $1$ &  & 8 \\
 10 & $B_3 + A_1$ & 1 & & 10 \\
 11 & $G_2+A_1$ & 1 & & 8\\
 12 & $G_2$ & 1 & & 4 \\
 13 & $B_3$ & 1 & & 5 \\
 14 & $2 A_1$  & $1$ & & 4 \\
 15 & $3 A_1$ & $1$ & & 8 \\
 16 & $3 A_1$ & $S_3$ &  
\begin{tikzpicture}[baseline=0,scale=.4,>=to]
\draw [<->] (0.22,0) to (1.78,0);
\draw [<->] (0.15,0.27) to (0.88,1.462);
\draw [<->] (1.2,1.462) to (1.8,0.27);
\filldraw (0,0) circle [radius=0.08];
\filldraw (2,0) circle [radius=0.08];
\filldraw (1,1.732) circle [radius=0.08];
\end{tikzpicture}
& 4  \\
 17 & $2 A_1$ & $1$ & & 4 \\
 18 & $C_3$ & 1 & & 4 \\
 19 & $2 A_1$ & $S_2$ & 
 \begin{tikzpicture}[baseline=0,scale=.4,>=to]
\draw [<->] (0.2,0) to (1.8,0);
\filldraw (0,0) circle [radius=0.08];
\filldraw (2,0) circle [radius=0.08];
\end{tikzpicture}
& 3 \\
 20 & $A_1+T_1$ & $S_2$ & & 0 \\
 21 & $A_2+T_1 $ & $S_2$ &  & 0 \\
 22 & $A_1$ & 1 & & 2\\
 23 & $G_2$ & 1 & & 4 \\
 24 &  $B_2$ & 1 & & 3\\
 25 & $T_2$ & $S_2$ & & 0 \\
 26 & $A_1 +T_1$ & $S_2$ & & 0 \\
 27 & $A_1$  & 1 & & 2 \\
 28 & $2 A_1 $ & 1 & & 4\\
 29 & $A_1$ & 1 & & 2 \\
 30 & $A_1$ & 1 & & 2 \\
 31 & $A_1$ & 1 & & 2\\
 32 & $A_1$ & $S_2$ & & 2 \\
 33 & $2 A_1$ & 1 & & 4 \\
 34 & $1$ & $S_3$ & & 1\\
 35  & $A_1$ & 1 & & 2\\
 36  & $A_1$ & 1 & & 2 \\
 37  & $A_1$ & 1 & & 2\\
 38  & $1$ & $S_2$ & & 1 \\
 39  & $A_1$ & 1 & & 2\\
 40  & $T_1$ & $S_2$ & & 0 \\
 41 & $A_1$ & 1 & & 2 \\
 42  & $1$ & $S_2$ & & 1 \\
 43  & $1$ & 1 & & 1 \\
 44  & $1$ & 1 & & 1\\
 45  & $1$ & 1 & & 1\\
 \hline
 \caption{The number of the orbit in the \cite{C} table, the unbroken gauge algebra,
the non-identity component of the gauge group, as well as the
action on the gauge algebra, when relevant. The final column is the tally of massive
vacua.}
\end{longtable}
\noindent
On the $A_1 \oplus A_1 \oplus A_1$ gauge algebra left unbroken by the triple associated to orbit
number 16, the  $S_3$  component group acts as permutations
of the summands. 
We therefore compute the number of the $2^3$ tensor product vacua which are invariant under $S_3$. The number 
is equal to
$4$, which is then the contribution to the supersymmetric index associated to orbit number 16. In the case of 
orbit number 19, the unbroken $\mathbb{Z}_2$ factor acts by exchanging the two gauge groups of type
$A_1$, and the number of vacua is 3. The final tally is
\begin{eqnarray}
I_{E_7} &=&
%18 + 10+ 2 \times 7 + 9+ 2 \times 4+6+4+0+2^3+ 2 \times 5+ 2 \times 4+
%4+5+2^2+2^3
%\nonumber \\
%& & +(4)+2^2+4+(3)+0+0+2+4+3+0+0+2+2^2+2+2
%\nonumber \\
%& & +2+2+2^2+1+2+2+2+1+2+0+2+1+1+1+1
%\nonumber \\
%&=& 
174.
\end{eqnarray}
%Orbit number 6 : outer automorphism of $A_5$ gauge algebra. (Does not act.)
%Orbit number 8 : massless. 

% I find $y^2+12 y+174$ here. 

\subsubsection*{The algebra $E_8$}
The algebra $E_8$ exhibits 70 nilpotent orbits, catalogued in table \ref{E8table}.
%\begin{table}[H]
%\centering
\begin{longtable}{|c|c|c|c|c|}
 \hline
 Number & Gauge Algebra & Component group  & Action & Massive Vacua \\
\hline
 \endhead
\hline
\endfoot
 70 & $1$ & 1 & &      1 \\
 \hline
 \caption{The orbit in the table of \cite{C} , the type of centralizer,
the component group, as well as its
action on the gauge algebra, when relevant. The number of massive vacua results.}
\endlastfoot
 1 & $E_8$ & 1 &  & 30 \\
 2 & $E_7$ & 1 &  & 18  \\
 3 & $B_6$ & 1 &   & 11 \\
 4 & $F_4 + A_1$ & 1 & & 18\\
 5 & $E_6$ & $S_2$ & & 12 \\
 6 & $C_4$ & 1 & & 5 \\
 7 & $A_5$ & $S_2$ & & 6 \\
 8 & $B_3 + A_1$ & 1 & & 10 \\
 9 & $B_5$ & 1 &  & 9 \\
 10 & $G_2+A_1$ & 1 &  & 8 \\
 11 & $2G_2$ & $S_2$ & 
 \begin{tikzpicture}[baseline=0,scale=.4,>=to]
\filldraw (0,0) circle [radius=0.12];
\filldraw (2,0) circle [radius=0.12];
\draw [-] (0,0) to (2,0);
\draw [-] (0,0.12) to (2,0.12);
\draw [-] (0,-0.12) to (2,-0.12);
\draw [-] (1.3,0.3) to (1.5,0);
\draw [-] (1.3,-0.3) to (1.5,0);
%\triplearrow{arrows={-Implies}}{(.1,0) -- (1.9,0)}
\filldraw (0,1) circle [radius=0.12];
\filldraw (2,1) circle [radius=0.12];
\draw [-] (0,1) to (2,1);
\draw [-] (0,1.12) to (2,1.12);
\draw [-] (0,1-0.12) to (2,1-0.12);
\draw [-] (1.3,1.3) to (1.5,1);
\draw [-] (1.3,1-0.3) to (1.5,1);
%\triplearrow{arrows={-Implies}}{(.1,1) -- (1.9,1)}
\draw[<->] (1,0.17) to (1,0.83);
\end{tikzpicture}
% $G_2 \leftrightarrow G_2$ 
 & 10 \\
 12 & $G_2 + A_1$ & 1  &  & 8 \\
 13 & $B_3 + A_1$ & 1 &  & 10 \\
 14 & $D_4$ & $S_3$ &   &  6 \\
 15 & $F_4$ & 1 &   &  9 \\
 16 & $B_2$ & 1 &  &  3 \\
 17 & $B_2 + A_1$ & 1 &   & 6 \\
 18 & $3A_1$ & $S_3$ & % 
 \begin{tikzpicture}[baseline=0,scale=.4,>=to]
\draw [<->] (0.22,0) to (1.78,0);
\draw [<->] (0.15,0.27) to (0.88,1.462);
\draw [<->] (1.2,1.462) to (1.8,0.27);
\filldraw (0,0) circle [radius=0.08];
\filldraw (2,0) circle [radius=0.08];
\filldraw (1,1.732) circle [radius=0.08];
\end{tikzpicture}
& 4 \\
 19 & $B_2 + T_1$ & $S_2$ &  & 0 \\
 20 & $A_4$ & $S_2$ &   &  5 \\
 21 & $2A_1$ & 1 &   &  4 \\
 22 & $C_3$ & 1 &   &  4 \\
 23 & $A_2$ & $S_2$ &   &  3 \\
 24 & $A_2 + T_1$ & $S_2$ &  & 0 \\
 25 & $B_2$ & 1 &   &  3 \\
 26 & $A_3$ & $S_2$ &   &  4 \\
 27 & $A_1 + T_1$ & $S_2$ &  & 0 \\
 28 & $2A_1$ & $1$ &   &  4 \\
 29 & $G_2 + A_1$ & 1 &   &  8  \\
 30 & $2A_1$ & $1$ &  &  4 \\
 31 & $A_1$ & 1 &  & 2  \\
 32 & $A_2$ & $S_2$ &   & 3 \\
 33 & $G_2$ & $S_2$ &   & 4 \\
 34 & $B_3$ & 1 &   &  5\\
 35 & $A_1$ & 1 &   & 2 \\
 % \hline
 %
 %\end{tabular}
%
%\end{table}
%\begin{table}[H]
%\centering
%\begin{tabular}{|c|c|c|c|c|}
%\hline
 %Number & Gauge Algebra & Component group  & Action & Massive Vacua \\
 %\hline
 36 & $2A_1$ & 1 & & 4 \\
 37 & $A_1$ & 1 & & 2 \\
 38 & $2A_1$ & $S_2$ & 
 \begin{tikzpicture}[baseline=0,scale=.4,>=to]
\draw [<->] (0.2,0) to (1.8,0);
\filldraw (0,0) circle [radius=0.08];
\filldraw (2,0) circle [radius=0.08];
\end{tikzpicture}
& 3\\
 39 & $A_1$ & $S_2$ && 2 \\
 40 & $A_1$ & $S_3$ & & 2 \\
 41 & $2A_1$ & $1$ & &  4 \\
 42 & $1$ & $S_5$ & & 1 \\
 43 & $2A_1$ & $1$ & & 4 \\
 44 & $2A_1$ & $S_2$ &
 \begin{tikzpicture}[baseline=0,scale=.4,>=to]
\draw [<->] (0.2,0) to (1.8,0);
\filldraw (0,0) circle [radius=0.08];
\filldraw (2,0) circle [radius=0.08];
\end{tikzpicture}
& 3  \\
 45 & $A_1$ & 1 & & 2 \\
 46 & $A_1$ & $S_2$ & & 2\\
 47 & $A_2$ & $S_2$ & & 3 \\
 48 & $T_1$ & $S_2$ & & 0 \\
 49 & $B_2$ & 1 & & 3 \\
 50 & $G_2$ & 1 & & 4 \\
 51 & $T_1$ & $S_2$ & & 0 \\
 52 & $A_1$ & 1 & & 2 \\
 53 & $T_1$ & $S_2$ & & 0 \\
 54 & $A_1$ & $S_2$ & & 2 \\
 55 & $1$ & $S_3$ & & 1 \\
 56 & $T_1$ & $S_2$ & & 0 \\
 57 & $A_1$ & 1 & & 2 \\
 58 & $A_1$ & 1 & & 2 \\
 59 & $1$ & $S_3$ & & 1\\
 60 & $A_1$ & 1 & & 2 \\
 61 & $1$ & $S_3$ & & 1 \\
 62 & $A_1$ & 1 & & 2 \\
 63 & $1$ & $S_2$ & & 1 \\
 64 & $1$ & $S_2$ & & 1 \\
 65 & $A_1$ & 1 & &    2 \\
 66 & $1$ & $S_2$ & & 1\\
 67 & $1$ & $S_2$ & & 1\\
 68 & $1$ & 1 & &      1\\
 69 & $1$ & 1 & &      1
\label{E8table}
\end{longtable}
\noindent
There are 70 orbits, among which orbits number 11, 18, 38 and 44 deserve special care.
In orbit 11, the algebra of the unbroken gauge group is $G_2 \oplus G_2$,
and the two summands are exchanged by the $S_2$ component group. Of the  $4 \times 4$ vacua of pure $\mathcal{N}=1$ with gauge group $G_2 \times G_2$,  10 are invariant under the component group.
In orbit 18, the three $su(2)$ algebras are exchanged by the $S_3$ group as in orbit 16 of $E_7$, leading to 4 vacua. In the case of orbits 38 and 44, two $su(2)$ algebras are exchanged by the $S_2$ component group, so both contribute 3 vacua. The census yields
\begin{eqnarray}
I_{E_8} 
%&=&
%30+18+11+2 \times 9+12+5+6+2 \times 5+9+2 \times 4+ (10)+2 \times 4+2 \times 5 + 6+9+3
%\nonumber \\
%& & +2 \times 3+
%(4)+0+5+4+4+3+0+3+4+0+4+2 \times 4+4+2+3+4+5+2+4+2+(3)
%\nonumber \\
%& & +2+2+4+1+4+(3)
%+2+2+3+0+3+4+0+2+0+2+1+0+2+2+1+
%2+1+2
%\nonumber \\
%& & +1+1+2+1+1+1+1+1 
%\nonumber \\
&=& 301 \, .
\end{eqnarray}
%I find $301 + 12 y$ for the index. 
We have thus computed all Witten indices of ${\cal N}=1^\ast$ supersymmetric
Yang-Mills theory on $\mathbb{R}^4$.

\section{Conclusions}
\label{conclusions}
We computed the number of massive vacua for ${\cal N}=1^\ast$ gauge theory on $\mathbb{R}^4$ for general
gauge group. The main technique we used was to find a bijection between the  classical vacuum expectation values for
the three massive adjoint chiral multiplets and the nilpotent orbits in the gauge algebra. Results on centralizers
of nilpotent orbits then allowed us to apply our knowledge of the supersymmetric indices of pure ${\cal N}=1$ super Yang-Mills theory to the
semi-classical solutions, thus providing us with a result for the tally of massive vacua. A subtlety resided
in the action of the component group of the centralizer on the summands of the unbroken gauge algebra.

There are many remaining problems in this realm. We already noted in the introduction that it will be
instructive to revisit the analysis of commuting triples in the light of the more recent remarks 
on the supersymmetric index for pure ${\cal N}=1$ on $\mathbb{R}^3 \times S^1$ \cite{Aharony:2013hda}. More closely related
to the present paper is the project of counting the number of massive vacua for ${\cal N}=1^\ast$ theory
on $\mathbb{R}^3 \times S^1$. 
The number of vacua for the $A$-type gauge groups turns out
to be  the same on $\mathbb{R}^4$ and on $\mathbb{R}^3 \times S^1$. For other types of gauge group, these
numbers can be different, for at least three reasons. The first is again the choice of global gauge
group and spectrum of line operators in the theory \cite{'tHooft:1981ht,Aharony:2013hda}. An example of this subtlety 
in the context of ${\cal N}=1^\ast$ theory was given in 
\cite{Bourget:2015cza}. The second reason is that massless vacua may allow for Wilson lines (taking values
in the gauge group) that commute with the adjoint vacuum expectation values, and that 
break the gauge group further to a non-abelian gauge group, providing further gapped vacua in the infrared
\cite{Bourget:2015cza}. The third reason is that Wilson lines can be turned on in the component group,
and this gives rise to multiple vacua in $\mathbb{R}^3 \times S^1$ originating from a single vacuum in
$\mathbb{R}^4$. We hope to return to the counting problem of vacua for ${\cal N}=1^\ast$ theory
on $\mathbb{R}^3 \times S^1$
 in the near future.

\section*{Acknowledgments}
We would like to thank  Amihay Hanany for useful discussions.
We would like to acknowledge support from the grant ANR-13-BS05-0001-02, and from the \'Ecole Polytechnique and the \'Ecole Nationale Sup\'erieure des Mines de Paris.

\appendix

\section{The Dual Coxeter Numbers}
\label{dualcoxeterapp}
We provide a table of dual Coxeter numbers for the simple complex Lie algebras.
\begin{table}[H]
\centering
\begin{tabular}{|c|c|c|c|c|c|c|c|c|
}
\hline 
$A_{n-1}$ & $B_n$ & $C_n$ & $D_n$ & $E_6$ & $E_7$ & $E_8$ & $F_4$ & $G_2$
\\
\hline
$n$ & $2n-1$ & $n+1$ & $2n-2$ & $12$ & $18$ & $30$ & $9$ & $4$
\\
\hline 
\end{tabular}
%\caption{
%The dual Coxeter numbers of simple Lie algebras}
%\label{dualcoxeternumbers}
\end{table}

\section{Illustrative Examples 
%ALT and Numerical Checks
}
\label{examples}
In this appendix, we illustrate some of the salient features of the analysis we presented in 
the bulk of the paper 
\label{Dexamples}
by two telling cases in which the special features of  $(S)O(4)$ (sub)groups
come into play.
\subsection{The gauge algebra $so(4)$}
In a first example, we discuss the ${\cal N}=1^\ast$ theories with $so(4)$ gauge algebra.
The $so(4)$ algebra is the direct sum of two $su(2)$ algebras. Thus, the theory 
with $SO(4)$ gauge group has the same vacuum structure as the theory with product gauge
group $SU(2) \times SU(2)$, and therefore has $3 \times 3$ massive vacua. Let us check this
elementary statement using our semi-classical method, based on the classification of nilpotent orbits:
\begin{equation}
\begin{array}{|c|c|c|}
\hline
\textrm{Orbit} & \textrm{Residual symmetry} & \textrm{Number of vacua} \\
\hline
 \{2,2\}_1 & Sp(2) & 2 \\
 \{2,2\}_2 & Sp(2) & 2 \\
 \{3,1\} & S(O(1) \times O(1)) & 1 \\
 \{1,1,1,1\} & SO(4) & 2 \times 2 \\
 \hline
\end{array}
\end{equation}
Yet another way to establish this result is by using the low energy effective superpotential on $\mathbb{R}^3 \times S^1$ (under the condition that no extra vacua arise upon compactification, which is satisfied in this case). This 
analysis has been done in 
\cite{Dorey:1999sj}. One can map out how the massive vacua behave under the infrared duality group \cite{Donagi:1995cf},
and one finds the tensor product of two duality diagrams of $SU(2)$ -- the latter is depicted in figure
\ref{dualities_su2}. The result is drawn in figure \ref{dualities_so4}.

When we consider the ${\cal N}=1^\ast$ theory with $O(4)$ gauge group, on the other hand, we have 
the table of nilpotent orbits and centralizers 
\begin{equation}
\begin{array}{|c|c|c|}
\hline
\textrm{Orbit} & \textrm{Residual symmetry} & \textrm{Number of vacua} \\
\hline
 \{2,2\} & Sp(2) & 2 \\
 \{3,1\} & O(1) \times O(1) & 1 \\
 \{1,1,1,1\} & O(4) & 2 \times 2 -1  \\
 \hline
\end{array}
\end{equation}
for a total of 6 vacua. The diagram of dualities is now the direct sum of two $SU(2)$ duality diagrams (i.e.
twice figure \ref{dualities_su2}).
We see the crucial role played by the $\mathbb{Z}_2$ identification of the two $su(2)$ 
summands in the Lie algebra when the gauge group is $O(4)$, as well as the fact that very even 
partitions correspond to a single orbit for $O(2n)$ groups.

\begin{figure}
\centering
\includegraphics[width=0.25\textwidth]{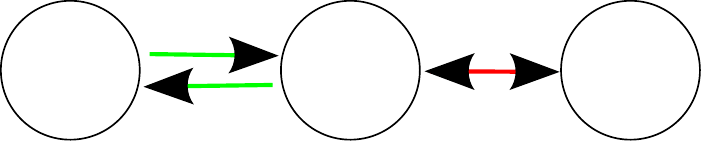}
\caption{The diagram of dualities for $SU(2)$. Each circle represents a massive vacuum. In red, we show the action of $S$-duality on the massive
vacua, in green $T$-duality (when non-trivial). }
\label{dualities_su2}
\end{figure}
\begin{figure}
\centering
\includegraphics[width=0.5\textwidth]{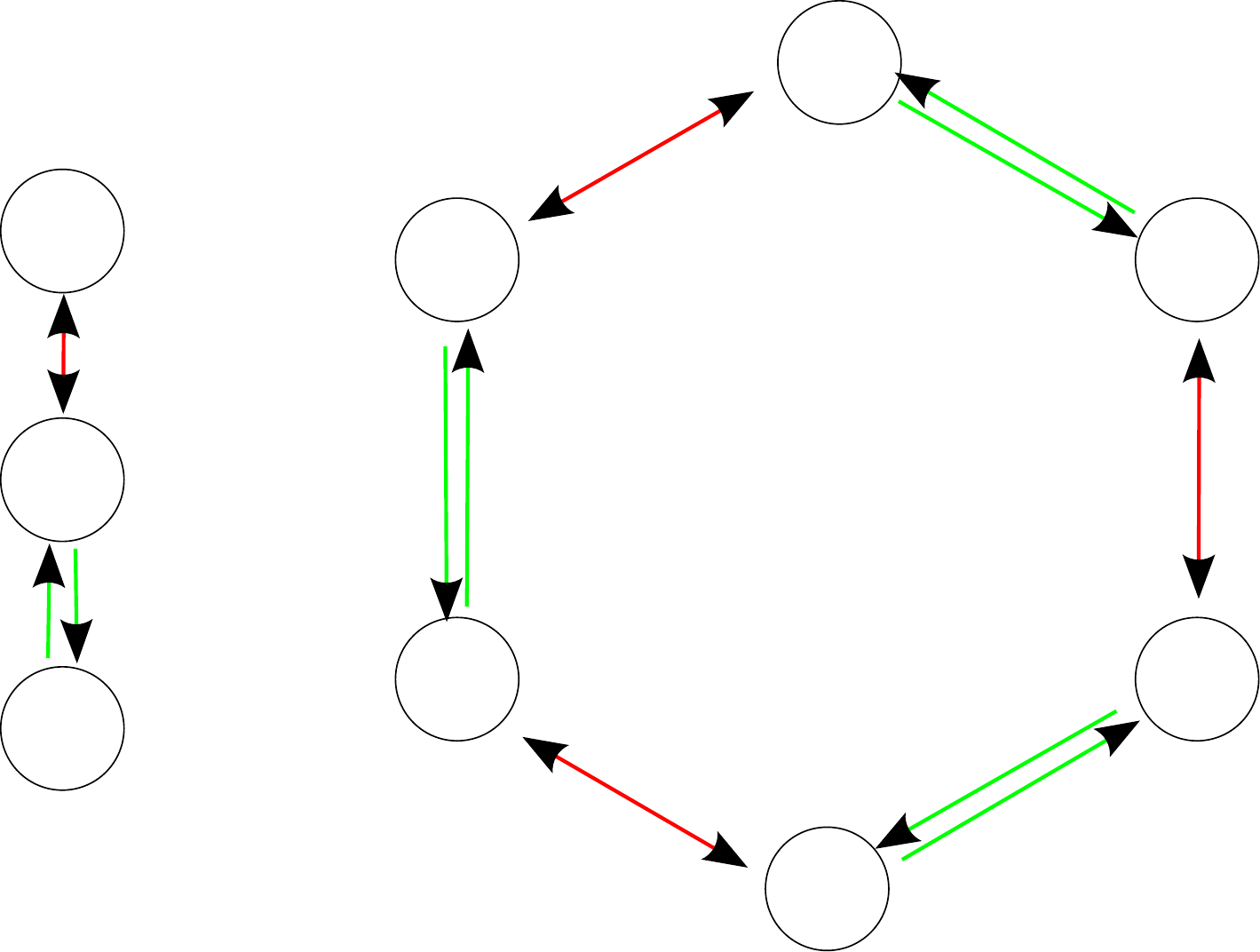}
\caption{The diagram of dualities for $SO(4)$. In red, we show the action of $S$-duality on the massive
vacua, in green $T$-duality (when non-trivial). }
\label{dualities_so4}
\end{figure}

\subsection{The gauge algebra $so(8)$}
Another interesting case that we wish to put forward is the case of the Lie algebra $so(8)$.
We note first of all that the group of outer automorphisms is the triality group $S_3$ that
acts on the $so(8)$ Dynkin diagram (in figure \ref{DynkinSO8}) by permuting the three external nodes.
In the following table, we list the partitions that correspond to the $SO(8)$ nilpotent orbits. We also
provide the Dynkin labels of the orbit (see \cite{C} for the necessary background). The latter labelling
is useful here, since it provides a direct handle on the behaviour of the orbits under the triality group.
\begin{equation}
\begin{array}{|c|c|c|c|c|}
\hline 
\textrm{Orbit} & \textrm{Dynkin diagram} & \textrm{Residual symmetry} & \textrm{Massive vacua} & \textrm{Triality Rep} \\
\hline
 \{4,4\}_1 & (0,2,0,2) & Sp(2) & 2 & 3_1 \\
 \{4,4\}_2 & (0,2,2,0) & Sp(2) & 2 & 3_1\\
 \{5,3\} & (2,0,2,2) & S(O(1)^2) & 1 & %\textrm{Subregular} 
 \\
 \{7,1\} & (2,2,2,2) &S(O(1)^2) & 1 & %\textrm{Principal} 
 \\
 \{2,2,2,2\}_1 & (0,0,0,2) & Sp(4) & 3 & 3_2 \\
 \{2,2,2,2\}_2 & (0,0,2,0) & Sp(4) & 3 & 3_2 \\
 \{3,2,2,1\} & (1,0,1,1) & Sp(2) \times S(O(1)^2) & 2 & \\
 \{3,3,1,1\} & (0,2,0,0) & S(O(2)^2) & 0 & \\
 \{5,1,1,1\} & (2,2,0,0) & S(O(3) \times O(1)) & 2 & 3_1 \\
 \{2,2,1,1,1,1\} & (0,1,0,0) & Sp(2) \times SO(4) & 8 & %\textrm{Minimal} 
 \\
 \{3,1,1,1,1,1\} & (2,0,0,0) & S(O(5) \times O(1)) & 3 & 3_2\\
 \{1,1,1,1,1,1,1,1\} & (0,0,0,0) & SO(8) & 6 & \\
 \hline
\end{array}
\end{equation}
The three orbits labelled $3_1$ in the last column form a triplet under triality, as do the three
orbits labelled $3_2$.
\begin{figure}
\centering
\includegraphics[width=0.25\textwidth]{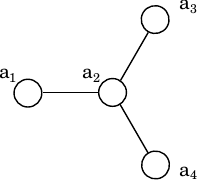}
\caption{This weighted Dynkin diagram for $so(8)$ is denoted $(a_1,a_2,a_3,a_4)$ in the text. }
\label{DynkinSO8}
\end{figure}
There are a total of 33 massive vacua for the ${\cal N}=1^\ast$ theory on $\mathbb{R}^4$ with gauge group
$SO(8)$. When we mod
out by a $\mathbb{Z}_2$ outer automorphism, which is equivalent to studying the gauge group $O(8)$,
we find $26$ massive vacua. To find this final tally, we need to realize that the $(0,1,0,0)$ orbit
contributes $2 \times (2 \times 2-1)=6$ vacua when the gauge group is  $O(8)$. Both these indices are in accord with the numerical 
analysis of the low-energy effective superpotential \cite{Kumar:2001iu} performed in \cite{Bourget:2015cza}, taking into account the
fact that one extra vacuum arises upon compactification on $S^1$ 
from the $(0,2,0,0)$ orbit.

\end{document}